\documentclass[journal]{IEEEtran}
\usepackage{amsmath}
\usepackage{amssymb}
\usepackage{amsthm}
\usepackage{bm}
\usepackage{bbm}
\usepackage{graphicx}
\usepackage{booktabs}
\usepackage[hyphens]{url}
\IfFileExists{cite.sty}{\usepackage{cite}}{}

\makeatletter
\g@addto@macro\UrlBreaks{%
\do\/\do-\do\.\do\_\do\?\do\&\do=\do\#\do\%\do\~}
\makeatother
\title{A New Analytical Method and Interpretation of Loss-of-Load Counts}

\author{
Aidan Looney,~\IEEEmembership{Graduate Student Member,~IEEE,}
Chanan Singh,~\IEEEmembership{Life Fellow,~IEEE,}
and
Le Xie,~\IEEEmembership{Fellow,~IEEE}%
\thanks{A. Looney and L. Xie are with the John A. Paulson School of
Engineering and Applied Sciences, Harvard University, Allston, MA, USA. C. Singh is with the Department of Electrical and Computer Engineering, Texas A\&M University, College Station, TX, USA. Corresponding author: Aidan Looney (aidanlooney@g.harvard.edu).}
}

\begin{document}

\maketitle

\begin{abstract}
This paper establishes the relationship between loss-of-load expectation (LOLE) and loss-of-load frequency (LOLF). We show why LOLE generally does not measure distinct-event frequency, mean event duration, or shortage magnitude. We identify conditions under which peak-based LOLE admits a frequency interpretation as the expected count of deficient daily peaks per horizon and distinguish this quantity from LOLF. We also derive an analytical formulation for frequency-and-duration indices that combines a convolution-derived generation probability-and-frequency model with chronological load changes. Under stationary generation, it uses the prescribed load profile directly, without constructing a separate load model or discretizing load magnitudes. Studies using generation and load data from the 24-bus IEEE RTS reveal different event frequencies and mean event durations at identical peak-based LOLE, motivating their separate assessment. Numerical comparisons demonstrate the accuracy advantage of the proposed analytical calculation over the conventional analytical implementations, quantifying the LOLF errors incurred by load discretization.
\end{abstract}

\begin{IEEEkeywords}
resource adequacy metrics, frequency-and-duration indices, loss-of-load expectation, loss-of-load frequency
\end{IEEEkeywords}

\section{Introduction}

Probabilistic methods for evaluating resource adequacy date to the 1930s, with daily-peak approaches emerging in the 1940s~\cite{billinton2015,rawg2022}. Loss-of-load expectation (LOLE) subsequently became a principal planning metric, associated with the widely used North American criterion of one day in ten years, or 0.1 days per year. Historically, this quantity represented the expected number of days on which available generation could not meet the daily peak load. However, counting days with insufficient capacity does not generally tell us how many distinct shortages occur (frequency), how long they last (duration), or the size of the outage (magnitude). 

Earlier literature contains interpretations of LOLE as a measure of shortage-event frequency. For example, Carden et al.\ described LOLE in 2011 as the expected number of firm load-shedding events over one or more years~\cite{carden2011_economics_ra_planning}. To address inconsistencies in LOLE's definition and interpretation, the IEEE Resource Adequacy Working Group (RAWG) published guidance in 2022 defining LOLE as the expected count of event-periods per horizon, where an event-period contains at least one shortfall~\cite{rawg2022}. This definition distinguishes affected reporting periods from distinct shortage events: one event can span several periods, and several events can occur within a single period. The expected count of distinct events is instead measured by loss-of-load frequency (LOLF), also termed loss-of-load events (LOLEV). Under this convention, LOLE does not generally measure frequency, duration, or magnitude. Additionally, this does not enable conversion between event-periods: 0.1 event-days per year or loss-of-load-days (LOLD) cannot be converted to 2.4 event-hours per year or loss-of-load hours (LOLH).

Despite this guidance, subsequent publications continue to describe LOLE as measuring event frequency. The Energy Systems Integration Group (ESIG)'s summary of its 2024 resource adequacy criteria report characterizes LOLE as measuring outage frequency~\cite{esig2024_new_ra_criteria_web}, while a 2024 Midcontinent Independent System Operator (MISO) presentation associates LOLE with event frequency and LOLH with event duration~\cite{miso2024_alternative_ra_criteria} even after citing the paper which clarified the interpretation of these metrics. These incorrect interpretations indicate a continuing need to distinguish affected-period counts, distinct-event frequency, and shortage duration. We therefore establish conditions under which peak-based LOLE admits a deficient-peak frequency interpretation and distinguish this quantity from the shortage-event frequency measured by LOLF.

The distinction between LOLE and LOLF also has consequences for their calculation. LOLE can be obtained by summing the probabilities of shortfall within reporting periods, whereas LOLF requires information about transitions into or out of shortage. Classical frequency-and-duration methods account for these transitions through changes in both generation availability and load~\cite{singh1975_frequency,billinton_allan2013_reliability,endrenyi1978_reliability_modeling}. One established approach represents chronological load transitions through a separate load model, which is combined with the generation model to construct a reserve-margin model~\cite{singh2019_grid_reliability}. Building on this framework, we evaluate load-induced shortage transitions directly along the prescribed load sequence, avoiding the intermediate construction of a separate load-transition model.

This paper addresses the relationship between LOLE and LOLF through both interpretation and calculation. The contributions are as follows:
\begin{enumerate}
    \item We clarify why LOLE does not generally measure outage frequency, duration, or magnitude, while showing that LOLH measures expected total shortage duration in hours when shortage status is constant within each hour.
    \item We establish conditions under which peak-based LOLE can be interpreted as a frequency of loss-of-peak, which differs from LOLF.  
    \item We derive a novel analytical formulation for computing LOLP and LOLF that combines generation transitions with the original chronological load values, avoiding load-level discretization errors.
\end{enumerate}

The rest of this paper is organized as follows. Section~\ref{sec:lole_interp} further clarifies the exact interpretation of LOLE and LOLF. Section~\ref{sec:freq_interp} provides the conditions under which peak-based LOLE can be interpreted as a deficient-peak frequency. Section~\ref{sec:lolf_analytic} then derives an analytical method for determining LOLF. Section~\ref{sec:numerical} then provides a numerical demonstration.

\section{Interpretation of Loss of Load Expectation}
\label{sec:lole_interp}

We adopt the event-period definition of LOLE proposed by the IEEE RAWG~\cite{rawg2022}. Let $[0,T)$ denote the study horizon, partitioned into reporting periods $B_1,\ldots,B_m$, and let $U(t)\geq 0$ denote unserved power. A reporting period is an event-period if a shortage occurs at any time within it. 

\begin{figure}[t]
    \centering
    \includegraphics[width=\columnwidth]
        {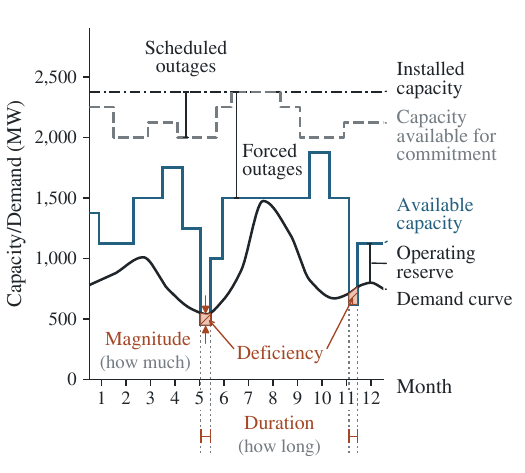}
    \caption{Depiction of the resource adequacy problem. Shortages occur when demand exceeds available capacity; their duration and magnitude are indicated and shaded regions represent unserved energy.}
    \label{fig:generation_load}
\end{figure}

Figure~\ref{fig:generation_load} illustrates the relationship between available generation capacity and demand. A shortage occurs when demand exceeds available capacity. Each uninterrupted shortage interval constitutes an event, its length gives the event duration, and the capacity deficit gives the instantaneous unserved power. LOLE instead counts reporting periods containing shortage, which need not correspond one-to-one with these events. Define $J_j=\mathbbm{1}\{\exists\,t\in B_j:U(t)>0\}$, where $\mathbbm{1}\{\cdot\}$ equals 1 when the condition holds and 0 otherwise. Let $N_{\mathrm{per}}=\sum_{j=1}^{m}J_j$ count the event-periods. Then
\begin{equation}
    \mathrm{LOLE}
    = \mathbb{E}[N_{\mathrm{per}}]
    = \sum_{j=1}^{m}\Pr(J_j=1).
    \label{eq:lole_period_count}
\end{equation}
Daily and hourly reporting periods give LOLD and LOLH, respectively. Both the reporting period and the study horizon must therefore be specified. A historical peak-based calculation instead evaluates deficiency at a selected peak-load observation within each period; it need not identify every period containing a shortage.

Distinct-event frequency concerns the number of separate shortage occurrences. At the finest temporal resolution represented by the model, an event is an uninterrupted episode of positive unserved power, with successive events separated by restored service. Let $N_E$ count the distinct shortage events occurring wholly or partly within the study horizon. Each event is counted once, including an event already in progress at the start of the horizon or still ongoing at its end. Using this counting convention, loss of load frequency (LOLF), also termed loss of load events (LOLEV), is
\begin{equation}
    \mathrm{LOLF}=\mathbb{E}[N_E],
    \label{eq:lolf_event_count}
\end{equation}
reported as expected events per horizon. LOLE and LOLF are therefore expectations of different counts. One event can extend across several reporting periods, while several events can occur within one reporting period.

\begin{figure}[t]
    \centering
    \includegraphics[width=\columnwidth]
        {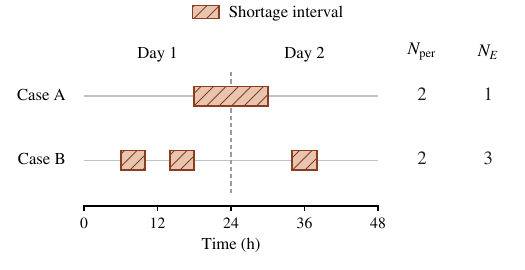}
    \caption{Counting distinction for two shortage cases. Both have $N_{\mathrm{per}}=2$ and 12 hours of shortage, but Case~A has $N_E=1$ and Case~B has $N_E=3$.}
    \label{fig:lole_event_count}
\end{figure}

Figure~\ref{fig:lole_event_count} illustrates this distinction using two shortage cases over a two-day horizon. Both cases affect two days and contain 12 hours of shortage. Case~A contains one event spanning the day boundary, whereas Case~B contains three separate events. Thus, the same affected-day count can correspond to different event counts, even when total shortage duration is unchanged. Consequently, a LOLE criterion of $0.1$ event-days per year does not generally imply $0.1$ distinct shortage events per year.

The assumption of constant shortage status within each hour is consistent with the hourly resolution commonly used in resource adequacy assessments, as acknowledged in~\cite{rawg2022}. When the generation--load balance is represented at this resolution, the model does not resolve the timing of shortage onset or recovery within an hour. Each deficient hour therefore contributes one hour of modeled shortage duration. Consequently,
\begin{equation}
    \mathrm{LOLH}=\mathrm{LOLF}\,\bar d,
\end{equation}
where $\bar d$ is the average duration of all shortage events in hours. Under this representation, LOLH has a duration interpretation as the expected aggregate shortage duration, while LOLF measures event frequency. Neither quantity alone determines the mean duration of an individual event.

The RAWG reference states that LOLH measures the expected count of hours containing a shortage, rather than the expected total shortage duration~\cite{rawg2022}. In a continuous-time interpretation, this distinction matters because a shortage may occupy only part of a counted hour. However, as the reference also acknowledges, resource adequacy assessments commonly use hourly temporal resolution. In an hourly model that represents shortage status as constant within each hour, each deficient hour contributes one full hour of shortage duration. Under this representation, LOLH does, in fact, measure expected total shortage duration.

LOLE also does not quantify the amount of unserved power or energy. Expected unserved energy, $\mathrm{EUE}=\mathbb{E}[\int_0^T U(t)\,dt]$, accounts for both the magnitude and duration of shortages. LOLE alone therefore generally does not determine event frequency, elapsed shortage duration, or shortage magnitude. The following section establishes the conditions under which peak-based LOLE admits a deficient-peak-frequency interpretation.

\section{Frequency Interpretation of Peak-Based Loss of Load Expectation}
\label{sec:freq_interp}
In this section, we identify conditions under which peak-based LOLE can have a frequency interpretation but acknowledge that this frequency is different from LOLF. Classical analytical frequency-and-duration methods use a state-space representation of generation and load~\cite{singh1975_frequency}. Let $A_t$ and $L_t$ denote available capacity and load at time $t$, respectively. The difference between the two states is therefore the margin: $M_t = A_t-L_t$. If $M_t \geq 0$, then the system has enough resources, but if $M_t < 0$, then the system is in a loss-of-load state. New shortage events begin when the system moves from a resource-adequate state to a loss-of-load state. Under stationary frequency balance, the entry and exit rates are equal. The system can exit the loss-of-load state through either an increase in available generation or a decrease in load. The component due to load could be calculated as the following where states $i,j$ are states for the available capacity and load, respectively such that $i-j < 0$:
\begin{align}
    \Pr(A_t=i, L_t=j)\lambda^L_{j\rightarrow k},
\end{align}
Here, $\lambda^L_{j\rightarrow k}$ is the load transition rate and $k\leq i$ is an adequate post-transition load level. Summing over all such $i,j,k$ gives the load-driven exit rate in events per unit time.

Let $PL_t$ be the daily peak load on day $t$, and let $A(i)$ denote the available capacity in generation state $i$. Assume that generation state $i$ has probability $p_i$ at every daily peak:  $p_i=\Pr(A_t=A(i))$. The daily peak-shortage probability is
\begin{align}
    \mathrm{LOLP}_{\mathrm{pk}}=\sum_i p_i\,\Pr(PL>A(i)).
\end{align}
The empirical peak exceedance probability is
\begin{align}
\Pr(PL>A(i))=\frac{\sum_{t=1}^{TP}\mathbbm{1}_{\{PL_t>A(i)\}}}{TP},
\end{align}
where $TP$ is the number of daily peaks in the horizon (365 for a typical calendar year). Peak-based LOLE is therefore
\begin{align}
\mathrm{LOLE}_{\mathrm{pk}}=TP\,\mathrm{LOLP}_{\mathrm{pk}}=\sum_i p_i\sum_{t=1}^{TP}\mathbbm{1}_{\{PL_t>A(i)\}}.
\label{eq:lole_pk_expression}
\end{align}

Substituting the empirical peak-exceedance probability into the peak-based LOLE expression cancels out the $TP$ factor. The inner sum counts the number of daily peaks that exceed each available-capacity level $A(i)$ during the horizon and therefore represents the deficient-peak frequency at that capacity level. Equation \eqref{eq:lole_pk_expression} consequently expresses peak-based LOLE as a probability-weighted sum of deficient-peak counts. This provides a frequency interpretation through the daily peak-exceedance counts, with generation availability represented by state probabilities rather than transition rates. Equation \eqref{eq:lole_pk_expression} therefore shows a frequency interpretation for peak-based LOLE, with the following qualifications:
\begin{enumerate}
    \item The load model considers one selected daily peak observation per day.
    \item This frequency uses generation-state probabilities but does not include the frequency component due to generation changes.
    \item This frequency counts deficient peaks (loss-of-peak), not distinct shortage events, and thus cannot generally replace LOLF when calculating mean event duration.
\end{enumerate}

\section{A New Analytical Technique for Frequency-and-Duration Metrics}
\label{sec:lolf_analytic}

This section derives an analytical method to calculate LOLP and LOLF directly from the original chronological load profile, avoiding errors caused by load-magnitude discretization. A recent NERC draft report \cite{nerc2026_probabilistic_adequacy_draft} states that LOLF cannot be calculated using analytical convolution methods because these methods do not include chronological information. We, however, derive such an analytical method for computing both LOLP and LOLF. 

Assume generating units fail and recover independently of one another, with these changes permitted at any time. The units are in steady state and the load profile has $N$ periods of length $\Delta =1$ hour. Let $C$ be installed system capacity, $L_t$ the load in period $t$, and $A_t$ the available generation capacity. For the set $\mathcal O$ of possible capacity outages, define $O_t=\min\{o\in\mathcal O:C-o<L_t\}$. The shortage probability in period $t$ is
\begin{align}
P_t=\Pr(A_t<L_t)=\Pr(A_t\leq C-O_t).
\label{eq:shortage_probability}
\end{align}
Each load period in the horizon has probability $1/N$. Weighting conditional shortage probabilities $P_t$ by the period probabilities gives LOLP. The expected shortage duration, denoted $D$, follows by multiplying the LOLP by the horizon length.
\begin{align}
    \mathrm{LOLP}=\frac{1}{N} \sum_{t=1}^N P_t, \quad   D = N \Delta \mathrm{LOLP}.
\end{align}
When shortage status is constant within each hour, $D=\mathrm{LOLH}$. For a year-long (8760-hour) horizon, $\mathrm{LOLH}=8760 \ \mathrm{LOLP}$. Evaluating $P_t$ at each original load value avoids errors caused by load-magnitude discretization.

By frequency balance~\cite{singh2019_grid_reliability}, the long-run rates of shortage entries and exits are equal. We calculate frequency by counting exits from shortage. Loss-of-load events can end through a change in the generation state or the load state, so both must be considered. The generation probability-and-frequency model is constructed once using the standard unit-addition procedure, which combines the individual unit models~\cite{singh2019_grid_reliability}. For each outage threshold $O_t$, this model provides $P_t$ and the stationary rate $F_t$ of exits from shortage, measured in events per hour. Let $\lambda^G_{i\to j}$ denote the mean transition rate from available capacity $i$ to $j$, conditional on capacity $i$. Then
\begin{equation}
    F_t=\sum_{i<L_t}\Pr(A_t=i)
        \sum_{j\geq L_t}\lambda^G_{i\to j}.
\end{equation}
Each term counts a generation change that restores sufficient capacity to meet demand. Changes between two shortage states do not end an event and therefore do not contribute to $F_t$. Generation changes consequently contribute $\Delta F_t$ expected exits in period $t$.

A decrease in load can also end a shortage. When $L_{t+1}<L_t$, this occurs if available generation cannot meet the previous load but can meet the new load. The probability of such a transition is
\begin{equation}
    \Pr(L_{t+1}\leq A_t<L_t)=P_t-P_{t+1}.
\end{equation}
Stationarity ensures that both probabilities use the same generation availability distribution. An increase in load cannot end a shortage, so the expected number of shortage exits at any load change is
\begin{equation}
    (P_t-P_{t+1})\mathbbm{1}_{\{P_t>P_{t+1}\}}.
    \label{eq:load_transition_exits}
\end{equation}
Combining the generation and load contributions using the period probabilities $1/N$ and load-transition rate $1/\Delta$ gives the shortage-exit frequency, in events per hour:
\begin{align}
    \mathrm{LOLF}
    &=\frac{1}{N}\sum_{t=1}^{N}F_t+\frac{1}{N\Delta}\sum_{t=1}^{N-1}
    (P_t-P_{t+1})\mathbbm{1}_{\{P_t>P_{t+1}\}}.
    \label{eq:analytical_lolf}
\end{align}
The load contribution includes the transitions between successive periods of the observed sequence. An event in progress at the end of the horizon is included in the event count of Section~\ref{sec:lole_interp} but not in the exit count. These counts differ by at most one and therefore yield the same long-run rate.

The connection between the two frequency interpretations can be understood through their underlying counts. Equation \eqref{eq:lole_pk_expression} weights the number of deficient daily peaks at each fixed capacity level by the probability of that capacity state. The load-transition term in \eqref{eq:analytical_lolf} similarly weights counts of downward load crossings that restore an adequate margin. These counts need not coincide: several daily peaks may lie within one uninterrupted interval above a capacity level, while several crossings may occur within a single day. The generation term additionally accounts for shortage exits caused by increases in available capacity. Thus, the distinction concerns both the occurrences counted and the transition information retained, explaining why the deficient-peak frequency cannot generally substitute for the event frequency used in frequency-and-duration analysis.

\section{Numerical Studies}
\label{sec:numerical}

The numerical studies examine how shortage-event frequency and mean duration vary at fixed peak-based LOLE. They also compare analytical frequencies with Monte Carlo estimates and quantify the errors avoided by our new analytical method.

\subsection{Test System and Study Parameters}

We use the generating-system data and load model of the IEEE Reliability Test System (RTS-79)~\cite{IEEE_RTS_1979}. The system comprises 32 generating units with 3405~MW of installed capacity and a peak demand of 2850~MW. Each unit is represented by an independent, stationary two-state continuous-time Markov process, with failure and repair rates $\lambda_g=1/\mathrm{MTTF}_g$ and $\mu_g=1/\mathrm{MTTR}_g$, respectively, using the published mean times to failure and repair. The 52-week load profile contains 8736 hours, or 364 days, and is periodically repeated. The assessment considers the basic resource adequacy problem without transmission constraints or scheduled maintenance.

Throughout this section, $\mathrm{LOLF}$ is reported as the expected number of shortage exits per 52-week profile, obtained by multiplying the hourly exit rate by $N\Delta = 8736$~h. The periodic implementation additionally includes the load transition between profiles, evaluated using \eqref{eq:load_transition_exits} with $P_{N+1} = P_1$. The analytical results are compared with event-driven sequential Monte Carlo simulation using $n=10^6$ independent profile realizations per case. Each realization starts with stationary unit states and counts shortage exits, including generation changes within hours and the closing load transition. 95\% confidence intervals for the Monte Carlo means use the normal approximation, with standard error $s/\sqrt{n}$, where $s$ is the sample standard deviation across realizations.

\subsection{Event Frequency at Fixed Peak-Based LOLE}

This study demonstrates that identical peak-based LOLE can correspond to different LOLF values and mean event durations. We vary generation transition rates while preserving the generation availability distribution, isolating information about shortage events that peak-based LOLE does not capture.

To separate outage probability from event frequency, we multiply both the failure and repair rates of every unit by $\alpha\in\{0.5,1,2\}$. This preserves each unit's stationary outage probability $\lambda_g/(\lambda_g+\mu_g)$ and hence the available-capacity distribution. Consequently, all three cases have the same peak-based index, $\mathrm{LOLE}_{\mathrm{pk}}=1.36886$ deficient peaks per profile, and the same expected total shortage duration, $D=\Delta\sum_t P_t=9.39418$~h per profile. The load-driven crossing contribution also remains unchanged, whereas the generation-driven contribution scales with $\alpha$. Numerically,
\begin{equation}
    \mathrm{LOLF}(\alpha)
    =0.347423\alpha+1.672252.
    \label{eq:numerical_rate_scaling}
\end{equation}

\begin{table}[t]
    \centering
    \caption{Event frequency per 52-week profile and analytical
    mean event duration.}
    \label{tab:numerical_frequency}
    \footnotesize
    \setlength{\tabcolsep}{2pt}
    \renewcommand{\arraystretch}{1.0}
    \setlength{\aboverulesep}{0.2ex}
    \setlength{\belowrulesep}{0.3ex}
    \begin{tabular*}{\columnwidth}{@{\extracolsep{\fill}}crrrc@{}}
        \toprule
        & \multicolumn{2}{c}{Analytical}
        & \multicolumn{2}{c}{Monte Carlo} \\
        \cmidrule(lr){2-3}\cmidrule(lr){4-5}
        $\alpha$
        & \multicolumn{1}{c}{LOLF}
        & \multicolumn{1}{c}{$\bar d$ (h)}
        & \multicolumn{1}{c}{Mean LOLF}
        & 95\% CI \\
        \midrule
        0.5 & 1.84596 & 5.089 & 1.84446 & [1.83793, 1.85099] \\
        1   & 2.01967 & 4.651 & 2.02333 & [2.01787, 2.02879] \\
        2   & 2.36710 & 3.969 & 2.36699 & [2.36210, 2.37188] \\
        \bottomrule
    \end{tabular*}
\end{table}

Table~\ref{tab:numerical_frequency} shows that increasing $\alpha$ from 0.5 to 2 raises LOLF by 28.23\%, although peak-based LOLE and total shortage duration are unchanged. The analytical long-run mean event duration, $\bar d=D/\mathrm{LOLF}$, decreases from 5.089 to 3.969~h. Additionally, the analytical LOLF lies within the Monte Carlo 95\% confidence interval in all cases, supporting the accuracy of the proposed calculation under the stated conditions. These results demonstrate that identical peak-based LOLE and expected total shortage duration can correspond to fewer, longer shortages or more frequent, shorter shortages, underscoring the need to assess event frequency and mean event duration separately.

\subsection{Effect of Load-Magnitude Discretization}
To quantify the LOLF errors introduced by load discretization, we compare the proposed calculation using the original load profile with a generation--load state-space implementation of conventional frequency-and-duration methods using discretized load states~\cite{singh2019_grid_reliability}.

\begin{figure}[t]
    \centering
    \includegraphics[width=0.85\columnwidth]{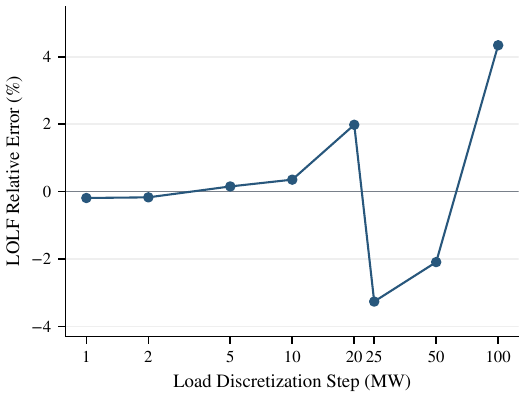}
    \caption{LOLF error introduced by discretizing load to a grid anchored at zero. This highlights the benefit of our method in eliminating these errors.}
    \label{fig:load_discretization}
\end{figure}

For the baseline case $\alpha=1$, we use load-grid widths $h$ of 1, 2, 5, 10, 20, 25, 50, and 100~MW. Each original hourly load is rounded to the nearest multiple of $h$, with ties rounded upward. Generation parameters and the chronological ordering of the load profile remain fixed. The discretized load bins are evaluated using the generation--load state-space reference. Figure~\ref{fig:load_discretization} reports the relative change from the original-load result of 2.01967470 events per profile.

The error is $-0.19\%$ at 1~MW, $-3.26\%$ at 25~MW, $-2.09\%$ at 50~MW, and $+4.34\%$ at 100~MW. Figure \ref{fig:load_discretization} demonstrates that the discretization error magnitude is non-monotonic in discretization size because rounding affects shortage probabilities and crossing frequencies when demand moves across attainable-capacity thresholds. Shifting the grid origin changes the rounded load values while keeping the grid width fixed. For example, shifting a 25~MW grid from levels $0,25,50,\ldots$ to $12.5,37.5,62.5,\ldots$~MW changes the LOLF error from $-3.26\%$ to $+4.96\%$. Thus, the LOLF error depends on both grid width and placement; our method avoids this sensitivity by retaining the original load magnitudes.

\section{Conclusion}
\label{sec:conclusion}

This paper addresses several critical issues in the common understanding of resource adequacy metrics. First, it clarifies that LOLE is the expected count of affected event-periods, while its peak-based form counts deficient peak observations; neither generally measures distinct-event frequency, mean event duration, or shortage magnitude. We then derive conditions under which peak-based LOLE can be interpreted as the frequency of loss-of-peak, but emphasize that this frequency counts deficient peaks rather than distinct shortage events. Finally, we derive a new analytical method for computing LOLP and LOLF under stationary generation and a prescribed hourly load profile without discretizing load magnitudes. The formulation accounts for generation- and load-driven shortage transitions directly without constructing the separate load-transition model used in conventional frequency-and-duration methods. The RTS-79 studies demonstrate the accuracy advantage of the proposed analytical LOLF calculation: by retaining the original chronological load values, it avoids LOLF errors of -3.26\% and +4.34\% induced by discretizing loads to 25 MW and 100 MW grids, respectively. Understanding the relationship between LOLE and LOLF is essential for interpreting resource adequacy criteria correctly, and this paper carries that distinction into an analytical evaluation of shortage events.

\section*{Disclaimer}
{The views expressed in this paper are the opinion of the
authors and do not reflect the views of PJM Interconnection,
L.L.C. or its Board of Managers, of which Le Xie is a member.}

\bibliographystyle{IEEEtran}
\bibliography{references}

\end{document}